\documentclass[prb,aps,twocolumn,showpacs,superscriptaddress,floatfix]{revtex4}
\usepackage{epsfig}
\usepackage{amsmath}
\usepackage{amssymb}
\usepackage{amsfonts}
\usepackage{mathptmx}
\usepackage{eucal}
\usepackage{bm}

\DeclareMathOperator{\re}{\mathop{\mathrm{Re}}}

\newcommand{\Eq}[1]{Eq.~(\ref{#1})}
\newcommand{\Eqs}[1]{Eqs.~(\ref{#1})}

\begin{document}

\title{Properties of tunnel Josephson junctions with a ferromagnetic interlayer}
\author{A.~S.~Vasenko}
\affiliation{Faculty of Science and Technology and MESA$^+$ Institute for Nanotechnology,
University of Twente, 7500 AE Enschede, The Netherlands}
\affiliation{Department of Physics, Moscow State University, Moscow 119992, Russia}
\author{A.~A.~Golubov}
\affiliation{Faculty of Science and Technology and MESA$^+$ Institute for Nanotechnology,
University of Twente, 7500 AE Enschede, The Netherlands}
\author{M.~Yu.~Kupriyanov}
\affiliation{Nuclear Physics Institute, Moscow State University, Moscow, 119992, Russia}
\author{M.~Weides}
\affiliation{Center of Nanoelectronic Systems for Information Technology (CNI), \\
Research Centre J\"{u}lich, D-52425 J\"{u}lich, Germany}
\date{\today}

\begin{abstract}
We investigate superconductor/insulator/ferromagnet/superconductor (SIFS)
tunnel Josephson junctions in the dirty limit, using the quasiclassical
theory. We formulate a quantitative model describing the oscillations of
critical current as a function of thickness of the ferromagnetic layer and
use this model to fit recent experimental data. We also calculate
quantitatively the density of states (DOS) in this type of junctions and
compare DOS oscillations with those of the critical current.
\end{abstract}

\maketitle


\section{Introduction}

It is well known that superconductivity and ferromagnetism are two
competing orders, however their interplay can be realized when the
two interactions are spatially separated. In this case the
coexistence of the two orderings is due to the proximity effect
\cite{RevB, RevG, RevV}. Experimentally this situation can be
realized in superconductor/ferromagnet (S/F) hybrid structures. The
main manifestation of the proximity effect in S/F structures is the
damped oscillatory behavior of the superconducting correlations in
the F layers. Two characteristic lengths of the decay and
oscillations are, correspondingly, $\xi _{f1}$ and $\xi _{f2}$.
Unusual proximity effect in S/F layered structures leads to a number
of striking phenomena like nonmonotonic dependence of their critical
temperature and oscillations of critical current in S/F/S Josephson
junctions upon the F layer thickness. Negative sign of the critical current
corresponds to the so-called $\pi$ state.
Spontaneous $\pi$ phase shifts in S/F/S junctions were observed
experimentally.\cite{Ryazanov, Kontos_exp, Blum, Sellier, Bauer,
Bell, Born, Shelukhin, Pepe, Oboznov, Weides, Weides2}

SIFS junctions, i.e. S/F/S trilayers with one transparent interface and one
tunnel barrier between S and F layers, represent practically interesting case of
$\pi$ junctions. SIFS structure offers the freedom to tune the critical
current density over a wide range and at the same time to realize high values
of a product of the junction critical current $I_{c}$ and its normal state resistance $R_{N}$.
\cite{Weides, Weides2} In addition, Nb based tunnel junctions are usually
underdamped, which is desired for many applications. SIFS $\pi $ junctions
have been proposed as potential logic elements in superconducting logic
circuits.\cite{logic} SIFS junctions are also interesting from the
fundamental point of view since they provide a convenient model system for a
comparative study between 0-$\pi$ transitions observed from the critical
current and from the density of states (DOS). At the same time, despite such
an interest, there is no complete theory yet of SIFS junctions which could
provide quantitative predictions for critical current and DOS in such
structures. All existing theories dealt only with a number of limiting cases,
when either linearized quasiclassical equations can be used for analysis
\cite{SIFS_lin} (e.g. temperature range near critical temperature, small
transparency of interfaces) or thickness of the F layer is small
\cite{RevB, RevG, RevV} compared to the decay characteristic length $\xi _{f1}$. Further,
in symmetric S/F/S junctions, the extension of theory to the case of nonhomogeneous
magnetization and large mean free path was performed in Refs.~\onlinecite{VolkovIc, RevV}.

The purpose of this work is to provide a quantitative model describing the
behavior of critical current and DOS in SIFS junctions as
a function of parameters characterizing material properties of the S, F
layers and the S/F interface transparency. The model provides a tool to fit
experimental data in existing SIFS junctions.

The paper is organized as follows. In the next section we formulate
the theoretical model and basic equations. In Sec.~\ref{Long} we
solve nonlinear Usadel equations, apply solutions for calculation of
critical current in SIFS junctions with long ferromagnetic layer, $d_f
\gg \xi_{f1}$, and fit recent experimental data. In
Sec.~\ref{General} we perform numerical calculations for critical
current in a SIFS junction with arbitrary length of the F layer. In
Sec.~\ref{DOS_FS} we numerically calculate DOS in the ferromagnetic
interlayer, and then summarize results in Sec.~\ref{Concl}.


\section{Model and basic equations}\label{Model}

\begin{figure}[tb]
\epsfxsize=8.5cm\epsffile{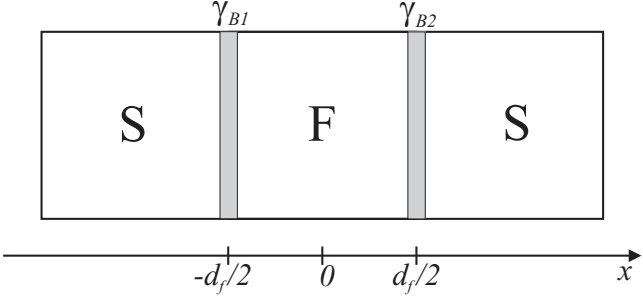} 
\caption{Geometry of the considered system. The thickness of the
ferromagnetic interlayer is $d_f$. The transparency of the left S/F
interface is characterized by the $\protect\gamma_{B1}$ coefficient, and the
transparency of the right F/S interface is characterized by $\protect\gamma_{B2}$.}
\label{SIFS}
\end{figure}

The model of an S/F/S junction we are going to study is depicted in
Fig.~\ref{SIFS} and consists of a ferromagnetic layer of thickness $d_{f}$ and two
thick superconducting electrodes along the $x$ direction. Left and right
superconductor/ferromagnet interfaces are characterized by the dimensionless
parameters $\gamma _{B1}$ and $\gamma _{B2}$, respectively, where $\gamma
_{B1,B2}=R_{B1,B2}\sigma _{n}/\xi _{n}$, $R_{B1,B2}$ are the
resistances of left  and right S/F interfaces, respectively, $\sigma _{n}$
is the conductivity of the F layer, $\xi _{n}=\sqrt{D_{f}/2\pi T_{c}}$,
$D_{f}$ is the diffusion coefficient in the ferromagnetic metal and $T_{c}$
is the critical temperature of the superconductor (we assume $\hbar =k_{B}=1$).
We also assume that the S/F interfaces are not magnetically active. We will
consider diffusive limit, in which the elastic scattering length $\ell $ is
much smaller than the decay characteristic length $\xi _{f1}$. In this paper
we concentrate on the case of a SIFS tunnel Josephson junction, when $\gamma
_{B1}\gg 1$ (tunnel barrier) and $\gamma _{B2}=0$ (fully transparent
interface). For comparison, we also consider two other limiting cases: an SFS
junction ($\gamma _{B1}=\gamma _{B2}=0$) and a SIFIS junction ($\gamma
_{B1},\gamma _{B2}\gg 1$).

Under conditions described above, the calculation of the Josephson
current requires solution of the one-dimensional Usadel
equations.\cite{Usadel} In the F layer the equations has the form
\cite{Pistolesi, Demler}
\begin{align}\label{Usadel_matrixF}
&D_f \frac{\partial}{\partial x}\left( \hat{G}_{f\uparrow(\downarrow)} \frac{\partial}{\partial x} \hat{G}_{f\uparrow(\downarrow)} \right)\nonumber
\\
&= \bigl[ (\omega \pm ih )\sigma_z + \frac{1}{2\tau_m}\sigma_z\hat{G}_{f\uparrow(\downarrow)}\sigma_z,
\hat{G}_{f\uparrow(\downarrow)} \bigr],
\end{align}
where positive sign ahead of $h$ corresponds to the spin up state $(\uparrow )$ and
negative sign to the spin down state $(\downarrow )$,
$\omega =2 \pi T(n + \frac{1}{2})$ are the Matsubara frequencies, $h$ is the
exchange field in the ferromagnet and
$\sigma_z$ is the Pauli matrix in the Nambu space.
The parameter $\tau_m$ is the spin-flip scattering time. The influence of spin-flip scattering
on various properties of S/F structures was considered in a number of papers.\cite{Oboznov, Volkov07,
Demler, Pistolesi, Faure, Gusakova, Cretinon}
We consider the ferromagnet with strong uniaxial anisotropy, in which case the magnetic
scattering does not couple the spin up and spin down electron populations.

The Usadel equation in the S layer can be written as\cite{Usadel}
\begin{equation}\label{Usadel_matrixS}
D_s \frac{\partial}{\partial x}\left( \hat{G}_s \frac{\partial}{\partial x} \hat{G}_s \right)
= \left[ \omega \sigma_z + \hat{\Delta}(x), \hat{G}_s \right],
\end{equation}
where $D_{s}$ is the diffusion coefficient in the superconductor.
In \Eq{Usadel_matrixS} $\hat{G}_s \equiv \hat{G}_{s\uparrow
(\downarrow)}$ and we omit subscripts `$\uparrow (\downarrow)$'
because equations in superconductor look identically for spin up
and spin down electron states.

In \Eqs{Usadel_matrixF}-\eqref{Usadel_matrixS} we use following
matrix notations (we omit `f', `s' and `$\uparrow(\downarrow)$'
subscripts)
\begin{equation}
\hat{G}(x, \omega) = \left(\begin{array}{cc} G & F \\
F^* & -G
\end{array}\right), \quad
\hat{\Delta}(x) = \left(\begin{array}{cc} 0 & \Delta(x) \\
\Delta^*(x) & 0
\end{array}\right),
\end{equation}
where $G$ and $F$ are normal and anomalous Green's functions, respectively,
and $\Delta(x)$ is the superconducting pair potential.
The matrix Green's function $\hat{G}$ satisfies the normalization condition,
\begin{equation}\label{norm}
\hat{G}^2 = 1, \quad G^2 + F F^* = 1,
\end{equation}
and the pair potential $\Delta(x)$ is determined by the self-consistency equation
\begin{equation}\label{s-c_matrix}
\Delta(x) \ln\frac{T_c}{T} = \pi T \sum_{\omega > 0} \left( \frac{2\Delta(x)}{\omega} - F_{s\uparrow}
-F_{s \downarrow} \right).
\end{equation}

The boundary conditions for the Usadel equations at the left and right sides
of each S/F interface are given by relations\cite{KL}
\begin{subequations}
\label{KL_matrix}
\begin{align}
\xi_n\gamma\left( \hat{G}_f \frac{\partial}{\partial x} \hat{G}_f \right)_{\pm d_f/2}
&= \xi_s \left( \hat{G}_s \frac{\partial}{\partial x} \hat{G}_s \right)_{\pm d_f/2},
\label{KL1_matrix} \\
2\xi_n \gamma_{B1} \left( \hat{G}_f \frac{\partial}{\partial x} \hat{G}_f
\right)_{-d_f/2} &= \left[ \hat{G}_s, \hat{G}_f \right]_{-d_f/2},
\label{KL_F_matrix} \\
2\xi_n \gamma_{B2} \left( \hat{G}_f \frac{\partial}{\partial x} \hat{G}_f
\right)_{d_f/2} &= \left[ \hat{G}_f, \hat{G}_s \right]_{d_f/2},
\label{KL_DOS_matrix}
\end{align}
\end{subequations}
where $\gamma = \xi_s\sigma_n/\xi_n\sigma_s$, $\sigma_s$ is the conductivity
of the S layer and $\xi_s = \sqrt{D_s/2\pi T_c}$.

To complete the boundary problem we also set boundary conditions at $x = \pm \infty$,
\begin{subequations}\label{gran}
\begin{equation}\label{granG}
G_s(\pm \infty) = \frac{\omega}{\sqrt{|\Delta|^2 + \omega^2}},
\end{equation}
\begin{equation}\label{granF}
F_s(-\infty) = \frac{|\Delta| e^{-i\varphi/2}}{\sqrt{|\Delta|^2 + \omega^2}}, \quad
F_s(+\infty) = \frac{|\Delta| e^{i\varphi/2}}{\sqrt{|\Delta|^2 + \omega^2}},
\end{equation}
\end{subequations}
where $\varphi$ is the superconducting phase difference between S electrodes.

In Matsubara technique it is convenient to parameterize the Green's function in the following way,
making use of the normalization condition, \Eq{norm},\cite{Zaikin}
\begin{equation}\label{parametrization}
\hat{G} = \left(\begin{array}{cc} \cos \theta & \sin \theta e^{i \chi} \\
\sin \theta e^{-i \chi} & -\cos\theta
\end{array}\right).
\end{equation}

Solving a system of nonlinear differential equations,
\Eqs{Usadel_matrixF}-\eqref{gran}, generally can be fulfilled only
numerically. We present full numerical calculation in
Sec.~\ref{General}. The analytical solution can be constructed in
case of one S/F bilayer, when we can set the phase $\chi$ in \Eq{parametrization} to zero.
We can also set the phase to zero in case of long S/F/S junction,
where the thickness of the ferromagnetic layer $d_f \gg \xi_{f1}$.
In that case, the decay of the Cooper pair wave function in
first approximation occurs independently near each interface.
Therefore we can consider the behavior of the anomalous Green's
function near each S/F interface, assuming that the ferromagnetic
interlayer is infinite. This analytical calculation for an S/F/S
trilayer with long ferromagnetic interlayer is performed in the next
section.

The general expression for the supercurrent is given by
\begin{equation}  \label{IC}
J_s = \frac{i \pi T \sigma_n}{4e} \sum\limits_{n=-\infty,\sigma=\uparrow,
\downarrow}^{+\infty} \left( \widetilde{F}_{f \sigma} \frac{\partial}{\partial x}
F_{f \sigma} - F_{f \sigma} \frac{\partial}{\partial x}
\widetilde{F}_{f \sigma} \right),
\end{equation}
where $\widetilde{F}_{f\uparrow(\downarrow)}(x, \omega) =
F^*_{f\uparrow(\downarrow)}(x, -\omega)$ are the anomalous Green's
functions in the ferromagnet.


\section{Critical current of junctions with long ferromagnetic interlayer}\label{Long}

We need to solve the complete nonlinear Usadel equations in the
ferromagnet, Eqs.~(\ref{Usadel_matrixF}). For SIFS junctions, an
analytical solution may be found if $d_{f}\gg \xi _{f1}$ and we can
set the phase of the anomalous Green's function to zero (see
discussion in Sec.~\ref{Model}).

Setting $\chi_s = \chi_f = 0$ we have the following $\theta$-parameterizations of
the normal and anomalous Green's functions, \Eq{parametrization}, $G=\cos \theta $ and $F=\sin \theta$.
In this case we can write \Eqs{Usadel_matrixF} in the F layer as
\begin{equation}\label{Usadel}
\frac{D_{f}}{2} \frac{\partial ^{2}\theta _{f\uparrow (\downarrow )}}{\partial x^{2}}
= \left( \omega \pm ih + \frac{\cos \theta_{f\uparrow
(\downarrow )}}{\tau _{m}}\right) \sin \theta_{f\uparrow (\downarrow )}.
\end{equation}
In the S layer the Usadel equation, \Eq{Usadel_matrixS}, may be now written as
\begin{equation}\label{Usadel_S}
\frac{D_s}{2} \frac{\partial^2 \theta_s}{\partial x^2} = \omega \sin \theta_s - \Delta(x) \cos \theta_s.
\end{equation}
The self-consistency equation in the S layer acquires the form
\begin{equation}
\Delta (x)\ln \frac{T_c}{T} = \pi T \sum\limits_{\omega > 0}
\left( \frac{2\Delta (x)}{\omega}-\sin \theta _{s \uparrow} - \sin \theta _{s \downarrow} \right).
\label{Delta}
\end{equation}

In the case of $\chi_s = \chi_f = 0$, the boundary conditions, \Eqs{KL_matrix},
for the functions $\theta_{f,s}$ at each S/F interface can be written as
\begin{subequations}
\label{KL}
\begin{align}
\xi_n\gamma\left( \frac{\partial \theta_f}{\partial x} \right)_{\pm d_f/2}
&= \xi_s \left( \frac{\partial \theta_s}{\partial x} \right)_{\pm d_f/2},
\label{KL1} \\
\xi_n \gamma_{B1} \left( \frac{\partial \theta_f}{\partial x}
\right)_{-d_f/2} &= \sin\left( \theta_f - \theta_s \right)_{-d_f/2},
\label{KL_F} \\
\xi_n \gamma_{B2} \left( \frac{\partial \theta_f}{\partial x}
\right)_{d_f/2} &= \sin\left( \theta_s - \theta_f \right)_{d_f/2}.
\label{KL_DOS}
\end{align}
\end{subequations}

The boundary conditions at $x = \pm \infty$ are
\begin{equation}\label{gran1}
\theta_s(\pm \infty) = \arctan\frac{|\Delta|}{\omega}.
\end{equation}

In the equation for the supercurrent, \Eq{IC}, the summation goes over all
Matsubara frequencies. It is possible to rewrite the sum only over positive Matsubara
frequencies due to the symmetry relation,
\begin{equation}  \label{sym}
\theta_{f(s)\uparrow}(\omega) = \theta_{f(s)\downarrow}(-\omega).
\end{equation}
In what follows, we will use only $\omega > 0$ in equations containing $\omega$.

For the left interface (tunnel barrier at $x = -d_f/2$), a first integral of
Eq.~(\ref{Usadel}) leads to
\begin{equation}  \label{Int}
\frac{\xi_f}{2} \frac{\partial \theta_f}{\partial x} = - q \sin \frac{\theta_f}{2}
\sqrt{1 - \epsilon^2 \sin^2 \frac{\theta_f}{2}},
\end{equation}
where $\xi_f = \sqrt{D_f/h}$ and the boundary condition $\theta_f(x
\rightarrow \infty) = 0$ has been used.
In \Eq{Int} we use the following notations
\begin{subequations}
\label{qe}
\begin{align}
q &= \sqrt{2/h}\sqrt{\omega \pm i h + 1/\tau_m},  \label{q} \\
\epsilon^2 &= \left(1/\tau_m\right) \left( \omega \pm i h +
1/\tau_m \right)^{-1}.  \label{e}
\end{align}
\end{subequations}
Here we again adopt convention that positive sign ahead of $h$ corresponds to the spin up
state $(\uparrow)$ and negative sign to the spin down state $(\downarrow)$. Here and below we
did not write spin labels `$\uparrow(\downarrow)$' explicitly but imply them everywhere they needed.

For the right interface ($x = d_f/2$), a first integral of
\Eq{Usadel} leads to a similar equation,
\begin{equation}  \label{Int1}
\frac{\xi_f}{2} \frac{\partial \theta_f}{\partial x} = q \sin \frac{\theta_f}{2}
\sqrt{1 - \epsilon^2 \sin^2 \frac{\theta_f}{2}}.
\end{equation}

Following Faure \textit{et al}.\cite{Faure} we integrate Eq.~(\ref{Int}),
which gives
\begin{equation}  \label{Int2}
\frac{\sqrt{1 - \epsilon^2 \sin^2\frac{\theta_f}{2}} - \cos\frac{\theta_f}{2}
}{\sqrt{1 - \epsilon^2 \sin^2\frac{\theta_f}{2}} + \cos\frac{\theta_f}{2}} =
g_1 \exp \left( -2q \frac{d_f/2 + x}{\xi_f} \right).
\end{equation}

The integration constant $g_1$ in Eq.~(\ref{Int2}) should be determined from
the boundary condition at the left S/F interface, Eq.~(\ref{KL_F}). Since we
consider the tunnel limit ($\gamma_{B1} \gg 1$), we can neglect small $\theta_f$
in the right hand side of Eq.~(\ref{KL_F}) and also assume,
neglecting the inverse proximity effect,
\begin{equation}  \label{arctan}
\theta_s(-d_f/2) = \arctan \frac{|\Delta|}{\omega}.
\end{equation}
Then \Eq{KL_F} becomes
\begin{equation}\label{G(n)}
\xi_n \gamma_{B1} \left( \frac{\partial \theta_f}{\partial x} \right)_{-d_f/2} = - G(n),
\quad  G(n) = \frac{|\Delta|}{\sqrt{\omega^2 + |\Delta|^2}}.
\end{equation}
From Eqs.~(\ref{Int}) and \eqref{G(n)} we obtain the boundary value of $\theta_f$ at $x = -d_f/2$
and substituting it into \Eq{Int2} we finally get
\begin{equation}  \label{gL}
g_1 = \frac{G^2(n)}{16 \gamma_{B1}^2}\frac{1-\epsilon^2}{q^2}
\left( \frac{\xi_f}{\xi_n} \right)^2.
\end{equation}

Linearizing Eq.~(\ref{Int2}), we can now obtain the anomalous
Green's function in the ferromagnetic layer of the SIF tunnel
junction with infinite F layer thickness. Similar formula for the FS
bilayer with a transparent interface ($\gamma_{B2}=0$) was
developed by Faure \textit{et al}.\cite{Faure} [to obtain it one
should integrate \Eq{Int1} and then linearize the resulting
equation]. The anomalous Green's function at the center of the F
layer in a SIFS junction may be taken as the superposition of the two
decaying functions, taking into account the phase difference in
each superconducting electrode,
\begin{align}  \label{theta}
\theta_f &= \frac{4}{\sqrt{1-\epsilon^2}} \biggl[\sqrt{g_1} \exp\left( -q
\frac{d_f/2 + x}{\xi_f} -i\frac{\varphi}{2}\right)  \notag \\
&+\sqrt{g_2}\exp\left( q\frac{x - d_f/2}{\xi_f} + i\frac{\varphi}{2}\right)
\biggr].
\end{align}
The expression for $g_2$ was obtained in Ref.~\onlinecite{Faure} for the
rigid boundary conditions at the transparent FS interface, $\theta_f(d_f/2)
= \arctan\left( |\Delta|/\omega \right)$ and reads
\begin{subequations}
\begin{align}
g_2 &= \frac{(1 - \epsilon^2) F^2(n)}{[ \sqrt{(1 - \epsilon^2) F^2(n) +1} +
1 ]^2}, \label{g2}\\
\quad F(n) &= \frac{|\Delta|}{\omega + \sqrt{\omega^2 +
|\Delta|^2}}. \label{gR}
\end{align}
\end{subequations}
Using the above solutions and Eqs.~(\ref{IC}), \eqref{sym} we arrive at sinusoidal current-phase
relation in a SIFS tunnel Josephson junction with the critical current
\begin{equation}  \label{ICSIFS}
I_c R_N = \frac{16 \pi T}{e} \re\left[\sum\limits_{n=0}^{\infty} \frac{G(n)
F(n) \exp\left( -q d_f/\xi_f \right)}{\sqrt{(1-\epsilon^2)F^2(n)+1}+1}\right].
\end{equation}
Here and below we fix positive sign in the definition of $q$,
$\epsilon^2$ in \Eqs{qe}: $q = \sqrt{2/h}\sqrt{\omega + i h +
1/\tau_m}$, $\epsilon^2 = \left(1/\tau_m\right) \left( \omega + i
h + 1/\tau_m \right)^{-1}$. It is possible since we already
performed summation over spin states and have to define now
spin-independent values. In \Eq{ICSIFS} and below $R_N$ is a full
resistance of an S/F/S trilayer, which include both interface
resistances of left and right interfaces and the resistance of the
ferromagnetic interlayer. In case of SIFS and SIFIS junctions the
F layer resistance can be neglected compared to large resistance
of the tunnel barrier.

\begin{figure}[t]
\epsfxsize=8.5cm\epsffile{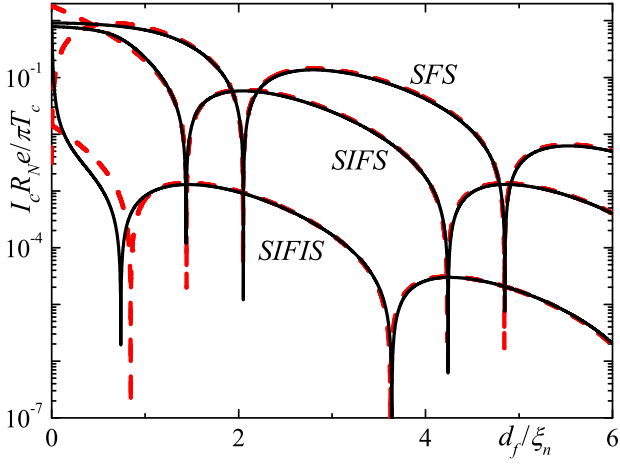} 
\caption{(Color online) The F layer thickness dependence of the
critical current for SFS ($\protect\gamma_{B1,2} = 0$), SIFS
($\protect\gamma_{B1} = 10^2$, $\protect\gamma_{B2} = 0$) and
SIFIS ($\protect\gamma_{B1,2} = 10^2$) junctions in the absence of
spin-flip scattering. Red dashed lines correspond to the modulus
of the analytical results \eqref{ICSFS},\eqref{ICSIFS} and
\eqref{ICSIFIS} and black solid lines correspond to the results of
numerical calculation in Sec.~\protect\ref{General}, $h = 3
\protect\pi T_c$, $T = 0.5T_c$.} \label{Compare} 
\end{figure}

At this point we define the characteristic lengths of the decay and oscillations $\xi_{f1,2}$ as,
\begin{subequations}
\begin{equation}
q/\xi_f = 1/\xi_{f1} + i/\xi_{f2},
\end{equation}
\begin{equation}  \label{xi12}
\frac{1}{\xi_{f1,2}} = \frac{1}{\xi_f}\sqrt{\sqrt{1 + \left( \frac{\omega}{h}
+ \frac{1}{h \tau_m} \right)^2} \pm \left( \frac{\omega}{h} + \frac{1}{h
\tau_m} \right)}.
\end{equation}
\end{subequations}

The critical current in Eq.~(\ref{ICSIFS}) is proportional to the
small exponent $\exp\left( -d_f/\xi_{f1} \right)$. The terms
neglected in our approach are of the order of $\exp\left( -2
d_f/\xi_{f1} \right)$ and they give a tiny second-harmonic term in
the current-phase relation.

The critical current equation \eqref{ICSIFS} can be simplified in
the limit of vanishing magnetic scattering, $\tau_m^{-1} \ll \pi
T_c$,
\begin{equation}  \label{ICSIFSa0}
I_c R_N = \frac{16 \pi T}{e}\sum\limits_{n=0}^{\infty} \left[\frac{G(n) F(n)
\exp\left( \frac{-d_f}{\xi_{f1}} \right) \cos\left( \frac{d_f}{\xi_{f2}}
\right)}{\sqrt{F^2(n)+1}+1} \right].
\end{equation}
Eq.~(\ref{ICSIFS}) also simplifies near $T_c$ and may be written as (for $T_c \ll h$)
\begin{equation}  \label{ICSIFSTc}
I_c R_N = \frac{\pi |\Delta|^2}{2 e T_c} \exp\left( -\frac{d_f}{\xi_{f1}}
\right) \cos\left( \frac{d_f}{\xi_{f2}}\right).
\end{equation}
The damped oscillatory behavior of the critical current can be clearly seen from
this equation. With increasing $d_f$ the junction undergoes the sequence of $0$-$\pi$
transitions when positive values of the $I_c R_N$ product
correspond to a zero state and negative values correspond to a $\pi$ state.

\begin{figure}[t]
\epsfxsize=8.5cm\epsffile{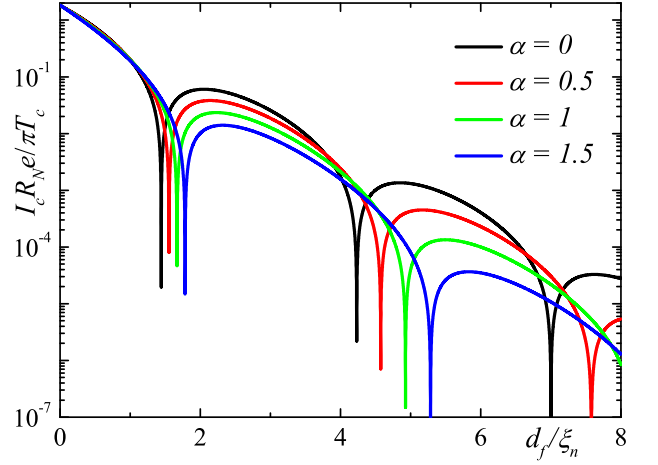} 
\caption{(Color online) The F layer thickness dependence of the critical
current in a SIFS junction [modulus of the Eq.~(\protect\ref{ICSIFS})] for
different values of $\protect\alpha = 1/ \protect\pi T_c \protect\tau_m$, $h
= 3 \protect\pi T_c$, $T = 0.5 T_c$.}
\label{diff_a}
\end{figure}

Eq.~(\ref{ICSIFSTc}) in the absence of spin-flip scattering
coincides with the corresponding equation, Eq.~(37), from the
Ref.~\onlinecite{SIFS_lin}, taken in the limit of long $d_f \gg
\xi_{f1}$ in case of $\gamma_{B1} \gg 1$, $\gamma_{B2} = 0$.

Using the same approach we can obtain the equation for the critical current
in a SIFIS structure with two strong tunnel barriers between the ferromagnet and
both superconducting layers ($\gamma_{B1,2} \gg 1$),
\begin{equation}  \label{ICSIFIS}
I_c R_N = \frac{4\pi T \xi_f}{e \xi_n} \frac{\gamma_{B1} + \gamma_{B2}}{\gamma_{B1}
\gamma_{B2}} \re\left[\sum\limits_{n=0}^{\infty} \frac{G^2(n)
\exp\left( \frac{-q d_f}{\xi_f} \right)}{q}\right].
\end{equation}
This formula coincides with corresponding expression Eq.~(39) for the critical
current in a SIFIS structure in Ref.~\onlinecite{Faure} for $\gamma_{B1,2} =
\gamma_B \gg 1$ and $d_f \gg \xi_{f1}$. Eq.~(\ref{ICSIFIS}) near $T_c$ may
be written as (for $T_c \ll h$)
\begin{align}  \label{ICSIFISTc}
I_c R_N = &\frac{\pi |\Delta|^2 \xi_{f2}}{2 e T_c \xi_n} \frac{\gamma_{B1} +
\gamma_{B2}}{\gamma_{B1} \gamma_{B2}}  \notag \\
&\times \cos\left( \Psi \right) \exp\left( \frac{-d_f}{\xi_{f1}} \right)
\sin \left( \Psi - \frac{d_f}{\xi_{f2}} \right),
\end{align}
where $\Psi$ is defined by $\tan\left( \Psi \right) = \xi_{f2}/\xi_{f1}$.
Eq.~(\ref{ICSIFISTc}) in the absence of spin-flip scattering coincides with
the corresponding equation, Eq.~(35), from the Ref.~\onlinecite{SIFS_lin},
taken in the limit of long $d_f \gg \xi_{f1}$.

\begin{figure}[tb]
\epsfxsize=8.5cm\epsffile{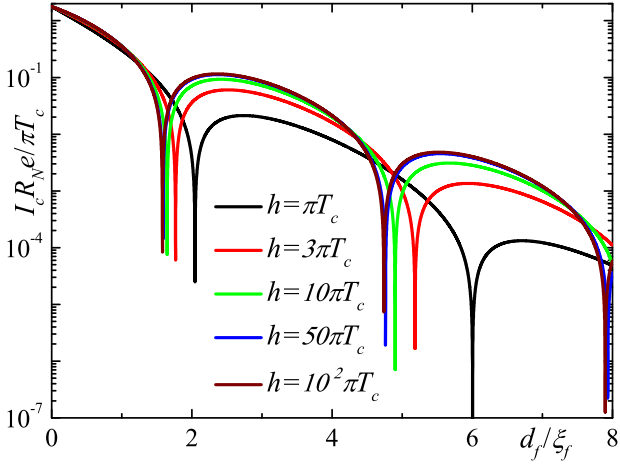} 
\caption{(Color online) The F layer thickness dependence of the critical
current in a SIFS junction [modulus of the Eq.~(\protect\ref{ICSIFS})] for
different values of exchange field $h$ in the absence of spin-flip
scattering, $T = 0.5 T_c$.}
\label{IC_h}
\end{figure}

We also provide here equation for the critical current in an SFS junction [see
Ref.~\onlinecite{Faure}, Eq.~(74)], written in our notations,
\begin{equation}  \label{ICSFS}
I_c R_N = \frac{64\pi T d_f}{e \xi_f} \re\left[\sum\limits_{n=0}^{\infty}
\frac{F^2(n) q \exp\left( -q d_f/\xi_f \right)}{[\sqrt{(1-\epsilon^2) F^2(n)
+ 1} + 1]^2}\right].
\end{equation}

We compare critical current dependencies over $d_{f}$ for SFS [Eq.~(\ref%
{ICSFS})], SIFS [Eq.~(\ref{ICSIFS})] and SIFIS [Eq.~(\ref{ICSIFIS})]
structures in Fig.~\ref{Compare}. Each of above junction types undergoes the
sequence of $0$-$\pi$ transitions with increasing thickness of the F layer.
From the figure we see that the transition
from $0$ to $\pi $ state occurs in SIFS tunnel junctions at shorter $d_{f}$
than in SFS junctions with transparent interfaces, but at longer $d_{f}$
than in SIFIS junctions with two strong tunnel barriers. This tendency can
be qualitatively explained by the fact that in structures with barriers
(SIFS, SIFIS) part of the $\pi $ phase shift occurs across the barriers.
Therefore a thinner F layer in a SIFS junction compared to an SFS one is needed
to provide the total shift of $\pi $ due to the order parameter oscillation. For
the same reason, $0$-$\pi $ transition in a SIFIS junction occurs at a smaller
thickness than in a SIFS junction. We note that in Fig.~\ref{Compare} we plot both
analytical and numerical calculated $I_c(d_f)$ dependencies, where numerical calculation
was performed for full boundary problem \Eqs{Usadel_matrixF}-\eqref{gran} [see further discussion
in Sec.~\ref{General}].

\begin{figure}[tb]
\epsfxsize=8.5cm\epsffile{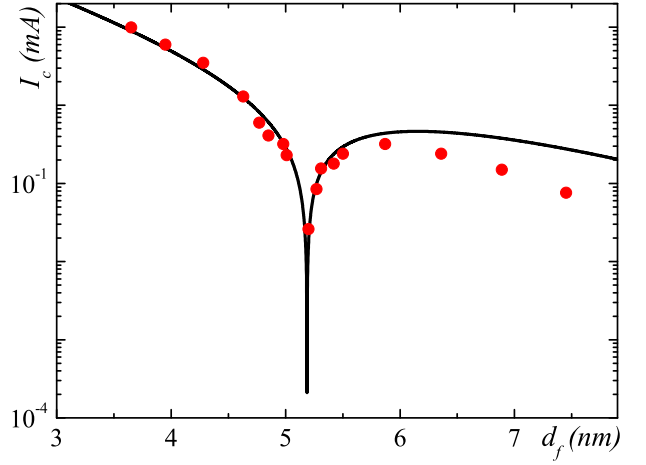} 
\caption{(Color online) Fit to the experimental data from
Ref.~\onlinecite{Weides} for the critical current in a
Nb/Al$_2$O$_3$/Ni$_{0.6}$Cu$_{0.4}$/Nb junction. The fitting parameters are:
$h/k_B = 950 \; K$ and $1/\protect\tau_m = 1.6 \; h$.}
\label{Weides}
\end{figure}

In Fig.~\ref{diff_a} we plot the F layer thickness dependence of the
critical current in a SIFS junction for different values of spin-flip
scattering time. For stronger spin-flip scattering the period of
supercurrent oscillations increases and the point of $0$-$\pi$ transition
shifts to the region of larger $d_f$. The same tendency exists for SFS and
SIFIS junctions.\cite{Faure}

In Fig.~\ref{IC_h} we plot the F layer thickness dependence of the
critical current in a SIFS junction for different values of the exchange
field $h$. We see that for large exchange fields $h\gg \pi T_{c}$ the
critical current scales with the ferromagnetic coherence length $\xi _{f}$.

From comparison with numerical results presented in Fig.~\ref{Compare} we
can conclude that the results for the critical current in SIFS junctions
presented in Figs.~\ref{diff_a}-\ref{IC_h} give correct magnitude of the
$I_c R_N$ product for $d_f \gtrsim \xi_n/2$.

As an application of the developed formalism, we present in Fig.~\ref{Weides}
the theoretical fit of the experimental data for a
Nb/Al$_2$O$_3$/Ni$_{0.6}$Cu$_{0.4}$/Nb junctions by
Weides \textit{et al}.\cite{Weides} making use of
Eq.~(\ref{ICSIFS}). We used following values of parameters: $R_B = 3.9 \;
m\Omega$, $D_f = 3.9 \; cm^2/s$, $T = 4.2 \; K$,\cite{Weides} $T_c = 7.2 \; K
$ (damped critical temperature in Nb). Good agreement was obtained with the
following parameters: $h/k_B = 950 \; K$, $1/\tau_m = 1.6 \; h$
(see Fig.~\ref{Weides}). These parameters can be compared with parameters obtained
by Oboznov \textit{et al}.\cite{Oboznov} for similar
ferromagnetic material, Ni$_{0.53}$Cu$_{0.47}$: $h/k_B = 850 \; K$, $%
1/\tau_m = 1.3 \; h$. Higher Ni concentration in the NiCu alloy in the experiment of
Weides \textit{et al}. results in higher exchange field.

In Ref.~\onlinecite{Oboznov} it was suggested that a ``dead'' layer
exists in the ferromagnet near each S/F interface, which does not take
part in the ``oscillating'' superconductivity. Other authors also
include into consideration the existence of nonmagnetic layers at
the interface of the ferromagnet and the superconductor or normal
metal.\cite{dl_1, dl_2, Cretinon} Thickness of the ``dead''
layer cannot be calculated quantitatively in the framework of our
model and also can not be directly estimated from the experiment. In
the experiment of Weides \textit{et al.}\cite{Weides} the range of
F layer thicknesses was rather narrow and only the first $0$-$\pi$
transition was observed. Due to these reasons we did not take into
account the existence of a nonmagnetic layer in our fit. This question
deserves separate detail experimental and theoretical study.

We should mention that the above estimates of exchange field and
spin-flip scattering time could be different if we consider
magnetically active S/F interfaces. It was shown in
Ref.~\onlinecite{Belzig} that the effect of spin-dependent boundary
conditions on the superconducting proximity effect in a diffusive
ferromagnet results in the change of the period of critical current
oscillations.


\section{Critical current of junctions with arbitrary length of the ferromagnetic
interlayer}\label{General}

\begin{figure}[t]
\epsfxsize=8.5cm\epsffile{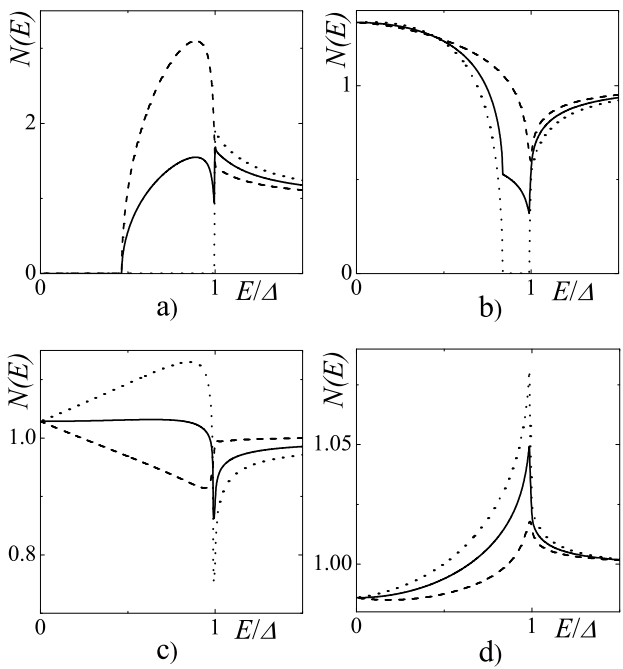} 
\caption{DOS on the free boundary of the F layer in the FS bilayer calculated
numerically in the absence of spin-flip scattering for different values of
the F layer thickness $d_f$: $N_\uparrow(E)$ (dashed line), $N_\downarrow(E)$
(dotted line) and $N(E)$ (solid line), $E_{ex} = 3 \protect\pi T_c$, $T = 0.5
T_c$. (a): $d_f/\protect\xi_n = 0.4$, (b): $d_f/\protect\xi_n = 1$, (c): $%
d_f/\protect\xi_n = 1.6$, (d): $d_f/\protect\xi_n = 2.2$.}
\label{DOS_a0}
\end{figure}
\begin{figure}[t]
\epsfxsize=8.5cm\epsffile{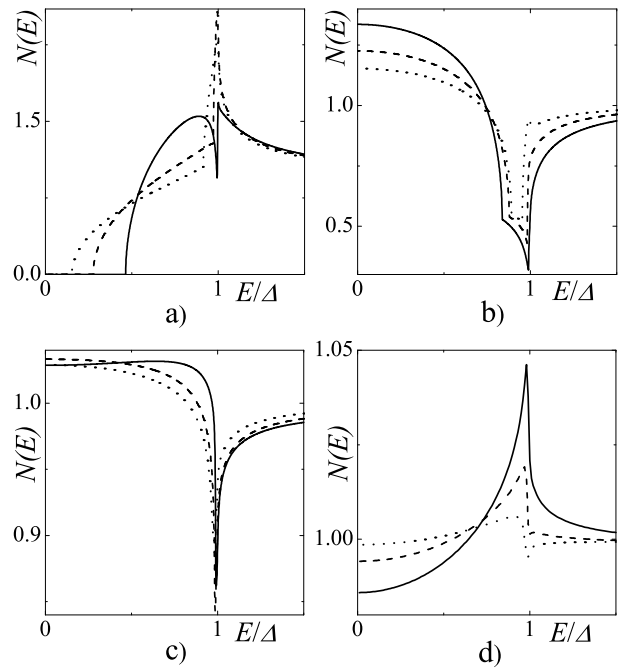} 
\caption{DOS $N(E)$ on the free boundary of the F layer in the FS bilayer
calculated numerically for $\protect\alpha = 1/\protect\pi T_c \protect\tau%
_m = 0$ (solid line), $\protect\alpha = 0.5$ (dashed line) and $\protect%
\alpha = 1$ (dotted line) for different values of the F layer thickness $d_f$%
, $E_{ex} = 3 \protect\pi T_c$, $T = 0.5 T_c$. (a): $d_f/\protect\xi_n = 0.4$%
, (b): $d_f/\protect\xi_n = 1$, (c): $d_f/\protect\xi_n = 1.6$, (d): $d_f/%
\protect\xi_n = 2.2$.}
\label{DOS_a}
\end{figure}

In the previous section we derived the expression for the critical current of
a SIFS junction in case of considerably long F layer thickness, $d_{f}\gg \xi
_{f1}$. For arbitrary F layer thickness in the absence of spin-flip
scattering, general boundary problem \eqref{Usadel_matrixF}-\eqref{gran} was solved
numerically using the iterative procedure.\cite{Golubov1} Starting from
trial values of the complex pair potential $\Delta(x)$ and the Green's functions
$\hat{G} _{s,f}$, we solve the resulting boundary problem.
After this we recalculate $\hat{G} _{s,f}$ and $\Delta(x)$. We
repeat the iterations until convergency is reached. The self-consistency of
calculations is checked by the condition of conservation of the supercurrent
across the junction.

In Fig.~\ref{Compare} we compare numerically and analytically
calculated $I_{c}(d_{f})$ dependencies in case of SFS, SIFS and
SIFIS junctions. We see that, as expected, the numerical method
provides correction only for small length of ferromagnetic layer. We
note that for SFS and SIFS junctions analytical curves \eqref{ICSFS}
and \eqref{ICSIFS} practically coincide with numerical results in
the region of the first $0$-$\pi$ transition.
For a SIFIS junction this transition occurs at smaller
$d_{f}$, where the assumptions of the section \ref{Long} are not
valid. However, in presence of strong spin-flip scattering the
first $0$-$\pi $ transition peak in a SIFIS junction shifts to the region of
larger $d_{f}$ and Eq.~(\ref{ICSIFIS}) describes the transition
accurately.

The main result of this section is that \Eq{ICSIFS} for the critical current of a SIFS junction
can be used as a tool to fit experimental data in SIFS junctions with good accuracy.


\section{Density of states oscillations in the ferromagnetic interlayer}\label{DOS_FS}

It is known that in a ferromagnetic metal attached to the superconductor the
quasiparticle DOS at energies close to the Fermi energy has a damped
oscillatory behavior.\cite{DOS1, DOS2, VolkovDOS} Experimental evidence for such
behavior was provided by Kontos \textit{et al}.\cite{Kontos} In SIFS
junctions we can compare the DOS oscillations with the critical current
oscillations.

We are interested in the quasiparticle DOS in the F layer in the
vicinity of the tunnel barrier ($x = -d_f/2 + 0$ in
Fig.~\ref{SIFS}). Below we will refer to the local DOS at this
point. For the case of strong tunnel barrier ($\gamma_{B1} \gg 1$)
left S layer and right FS bilayer in Fig.~\ref{SIFS} are uncoupled.
Therefore we need to calculate the DOS in the FS bilayer at the free
boundary of the ferromagnet. Solving numerically
Eqs.~(\ref{Usadel})-\eqref{gran1}, we set to zero the $\theta_f$
derivative at the free edge of the FS bilayer, $x = -d_f/2$, $\left(
\partial \theta_f/\partial x \right)_{-d_f/2} = 0$.\cite{Gusakova}

\begin{figure}[t]
\epsfxsize=8.5cm\epsffile{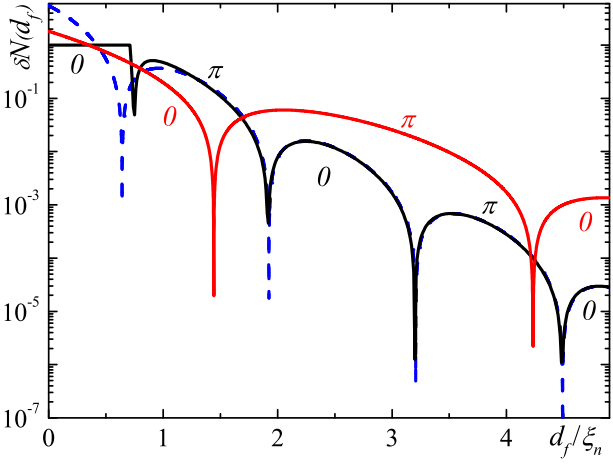} 
\caption{(Color online) The F-layer dependence of the function
$\protect \delta N(d_f)$ in the absence of spin-flip scattering, $h
= 3\protect\pi T_c$ , $T = 0.5 T_c$. Black solid line is a result of
the numerical calculation; blue dashed line is calculated with the
use of Eq.~(\protect\ref{r0}). Red line shows normalized critical
current for a SIFS junction. Zero and $\pi$ states defined from $I_c$
are indicated by red color, while zero and $\pi$ states defined from
the DOS are indicated by black color.} \label{DOS_zero} 
\end{figure}

We use the self-consistent two step iterative
procedure\cite{Golubov1,GK1,Gusakova}. In the first step we
calculate the pair potential coordinate dependence $\Delta(x)$ using
the self-consistency equation in the S layer, Eq.~(\ref{Delta}).
Then, by proceeding to the analytical continuation in
Eqs.~(\ref{Usadel}),~\eqref{Usadel_S} over the quasiparticle energy
$i\omega \rightarrow E + i0$ and using the $\Delta(x)$ dependence
obtained in the previous step, we find the Green's functions by
repeating the iterations until convergency is reached. We define
the full DOS $N(E)$ and the spin resolved DOS
$N_{\uparrow(\downarrow)}(E)$, normalized to the DOS in the normal
state, as
\begin{subequations}
\label{DOS}
\begin{align}
N(E) &= \left[ N_\uparrow(E) + N_\downarrow(E)\right]/2,  \label{DOS_full} \\
N_{\uparrow(\downarrow)}(E) &= \re\left[\cos\theta_{\uparrow(\downarrow)}(%
i\omega \rightarrow E + i0)\right].  \label{DOS_spin}
\end{align}

The numerically obtained energy dependencies of the DOS at the free F
boundary of the FS bilayer are presented in Figs.~\ref{DOS_a0} and \ref{DOS_a}.
Fig.~\ref{DOS_a0} demonstrates the DOS energy dependence for different $d_f$. At
small $d_f$ full DOS turns to zero inside a minigap, which vanishes with the
increase of $d_f$. Then the DOS at the Fermi energy $N(0)$ rapidly
increases to the values larger than unity and with further increase of $d_f$
it oscillates around unity while it's absolute value exponentially approaches
unity (see also Fig.~\ref{DOS_zero}). In Fig.~\ref{DOS_a0} we also plot
the spin resolved DOS energy dependencies $N_\uparrow(E)$ and $N_\downarrow(E)$.
Fig.~\ref{DOS_a} demonstrates full DOS energy dependence for different
values of spin-flip scattering time. For stronger spin-flip scattering the
minigap closes at smaller $d_f$, the period of the DOS oscillations at the Fermi
energy increases and the damped exponential decay occurs faster.

In case of long F layer ($d_f \gg \xi_{f1}$) it is also possible to obtain an
analytical expression for the DOS at the free boundary of the ferromagnet,
\end{subequations}
\begin{equation}  \label{DOS_bound}
N_{\uparrow(\downarrow)}(E) = \re[ \cos\theta_{b \uparrow(\downarrow)} ] \approx 1 - \frac{1}{2}\re \theta_{b \uparrow(\downarrow)}^2,
\end{equation}
where $\theta_{b \uparrow(\downarrow)}$ is a boundary value of $\theta_f$ at $x=-d_f/2$. It can be
obtained by the mapping method, similar to the one used in the electrostatic problems. We
consider the FS bilayer where $x \in [-d_f/2, \;
d_f/2]$ stands for the ferromagnetic metal and $x > d_f/2$ stands for the
superconductor; the point $x = - d_f/2$ corresponds to the free F
layer boundary. For infinite F layer ($d_f \rightarrow \infty$)
the solution for $\theta_{f \uparrow(\downarrow)}$ far from the interface is given by the
exponential term in Eq.~(\ref{theta}), written in the real energy
space,
\begin{equation}\label{theta_left}
\overleftarrow{\theta}_{f \uparrow(\downarrow)} =
\frac{4}{\sqrt{1-\eta^2}} \sqrt{g_2} \exp\left( p \frac{x -
d_f/2}{\xi_f} \right),
\end{equation}
where
\begin{subequations}
\label{qef}
\begin{align}
p &= \sqrt{2/h}\sqrt{-iE_R \pm ih + 1/\tau_m}, \label{p}\\
\eta^2 &= (1/\tau_m)(-iE_R \pm ih + 1/\tau_m)^{-1}, \label{eta}\\
g_2 &= \frac{(1 - \eta^2) F^2(E)}{[\sqrt{(1-\eta^2)F^2(E)+1} + 1]^2}, \\
F(E) &= \frac{|\Delta|}{-iE_R + \sqrt{|\Delta|^2 - E_R^2}}, \quad E_R = E + i0.
\end{align}
\end{subequations}
Here, as above, positive sign ahead of $h$ corresponds to the spin up state in \Eq{theta_left}
and negative sign for the spin down state.
By using the arrow `from right to left' in $\overleftarrow{\theta}_{f \uparrow(\downarrow)}$ we want to
stress that this solution is induced in the ferromagnet from the right FS interface.

In the case of finite ferromagnet length the
boundary conditions at the free F layer boundary, $x = -d_f/2$, become
\begin{equation}
\theta_{f \uparrow(\downarrow)}(-d_f/2) = \theta_{b \uparrow(\downarrow)}, \quad
\left( \frac{\partial \theta_{f \uparrow(\downarrow)}}{\partial x} \right)_{-d_f/2} = 0.
\end{equation}
To ensure these conditions we add another exponential solution,
\begin{equation}
\overrightarrow{\theta}_{f \uparrow(\downarrow)} =
\frac{4}{\sqrt{1-\eta^2}} \sqrt{g_2} \exp\left(
-p\frac{3d_f/2+x}{\xi_f} \right),
\end{equation}
resulting from the mirror image of
the F layer with respect to the point $x = -d_f/2$. At $x =
-d_f/2$ both exponential terms are equal to each other and the
final solution, $\theta_{b \uparrow(\downarrow)} = \overleftarrow{\theta}_{f \uparrow(\downarrow)}(-d_f/2) +
\overrightarrow{\theta}_{f \uparrow(\downarrow)}(-d_f/2)$, is two
times larger than the solution for infinite ferromagnetic
layer at this point and reads
\begin{equation}  \label{theta_bound}
\theta_{b \uparrow(\downarrow)}= \frac{8 F(E)}{\sqrt{(1-\eta^2)F^2(E) + 1} + 1} \exp\left(
-p\frac{d_f}{\xi_f}\right).
\end{equation}
This equation coincides with the result obtained in Ref.~\onlinecite{Cretinon}
by direct integration of the Usadel equation.

In Fig.~\ref{DOS_zero} we plot analytically and numerically calculated function
\begin{equation}  \label{r}
\delta N(d_f) = |1 - N_0|, \quad N_0 = N(E = 0),
\end{equation}
together with the $I_c(d_f)$ dependence for a SIFS junction. We see that the point of
$0$-$\pi $ transition on the $I_c(d_f)$ plot does not coincide with
the first minimum of $\delta N(d_f)$ corresponding to sign change of
$1 - N_0$. This difference can be qualitatively explained as
follows. The transition from $0$ to $\pi$ state in a junction, seen
as sign change of $I_c(d_f)$, is the result of interference of
solutions for $\theta_f$ originating from two S electrodes. $0$-$\pi$
transition in  $I_c(d_f)$ occurs approximately at such thickness
$d_f$ when the boundary value of $\theta_f$ in Eq.~(\ref{theta}) at
$x=-d_f/2$ becomes negative, i.e. when $\theta_f$ acquires the phase
shift $\pi$. On the other hand, sign change of $1 - N_0$ occurs at
such $d_f$ when the boundary value $\theta_b$ in
Eq.~(\ref{theta_bound}) becomes an imaginary number, i.e. when
$\theta_f$ acquires the phase shift $\pi/2$. It occures at smaller
$d_f$ compared to  $0$-$\pi$ transition in the critical current.
Corresponding $0$ and $\pi$ states defined from $I_c$ and from the DOS
are indicated in Fig.~\ref{DOS_zero}.

It is also seen from Fig.~\ref{DOS_zero} that the DOS oscillations have the period
approximately twice smaller than those of the critical current. This fact is easy to see
from the analytical expression for $\delta N(d_f)$. Using Eqs.~(\ref{DOS})-
\eqref{r} we obtain
\begin{equation}  \label{r_analyt}
\delta N(d_f) = 32\biggl|\re\biggl[\frac{1}{(\sqrt{2 -
\eta^2_0} + 1)^2} \exp\left( -p_0\frac{2 d_f}{\xi_f}\right)
\biggr]\biggr|,
\end{equation}
where $\eta_0 = \eta(E = 0)$ and $p_0 = p(E = 0)$ in
Eqs.~(\ref{p})-\eqref{eta}. At vanishing magnetic scattering,
$\tau_m^{-1} \ll \pi T_C$, this equation can be simplified,
\begin{equation}  \label{r0}
\delta N(d_f) = \frac{32}{3 + 2\sqrt{2}} \left|
\exp\left(-\frac{2 d_f}{\xi_{f1}}\right) \cos\left(\frac{2 d_f}{\xi_{f2}}\right) \right|,
\end{equation}
where characteristic lengths of decay and oscillations $\xi_{f1,2}$ are
given by Eq.~(\ref{xi12}) with the substitution $i\omega \rightarrow E + i0$.
This equation can be compared with Eq.~(\ref{ICSIFSa0}). We see that the
period of the DOS oscillations is approximately twice smaller than the period of
the critical current oscillations and the exponential decay is approximately twice
faster than the decay of the critical current.


\section{Conclusion}\label{Concl}

We have developed a quantitative model, which describes the oscillations of
the critical current as a function of the F layer thickness in a SIFS tunnel
junctions with thick ferromagnetic interlayer, $d_f \gg \xi_{f1}$, in the
dirty limit. We justified this model by numerical calculations in general
case of arbitrary $d_f$: for all values of parameters characterizing
material properties of the ferromagnetic metal numerical and analytical
results coincide in physically important region of the first $0$-$\pi$ transition.
Thus the derived analytical expression for the critical current can be used as a tool to fit experimental
data in various types of SIFS junctions. We have discussed the details of the
damped oscillatory behavior of the critical current for different values of
the F layer parameters.

We also studied the superconducting DOS induced in a ferromagnet by the
proximity effect. We showed that the oscillation pattern of DOS at the Fermi
energy in the ferromagnet (at location of the tunnel junction) does not coincide
with that of the critical current in a SIFS junction and it's period is
approximately twice smaller. Therefore the DOS oscillations do not reflect the
$0$-$\pi$ transition in $I_c(d_f)$. We calculated the quasiparticle DOS in the F layer in the close
vicinity of the tunnel barrier which can be used to obtain
current-voltage characteristics for a SIFS junction. These calculations will
be presented elsewhere.

Finally, we used our results to fit recent experimental data for
SIFS tunnel junctions and extracted important parameters of the
ferromagnetic interlayer.


\begin{acknowledgments}
The authors thank W.~Belzig, E.~V.~Bezuglyi, A.~I.~Buzdin, T. Champel, S.~Kawabata
and F. Pistolesi for useful discussions. This work was supported by NanoNed program
under Project No. TCS7029 and RFBR project N08-02-90012.
\end{acknowledgments}

\end{document}